\documentclass[twocolumn,showpacs,preprintnumbers,amsmath,amssymb,nofootinbib,superscriptaddress]{revtex4}

\usepackage{graphicx}
\usepackage{dcolumn}
\usepackage{bm}

\usepackage[usenames]{color} 



\def\be{\begin{equation}}
\def\ee{\end{equation}}
\def\bea{\begin{eqnarray}}
\def\eea{\end{eqnarray}}

\def\cmm2{{\,\rm cm^{-2}}}
\def\cm2{{\,{\rm cm}^2}}
\def\cmm3{{\,{\rm cm}^{-3}}}
\def\gcmm3{{\,{\rm g\,cm^{-3}}}}

\def\fun#1#2{\lower3.6pt\vbox{\baselineskip0pt\lineskip.9pt
  \ialign{$\mathsurround=0pt#1\hfil##\hfil$\crcr#2\crcr\sim\crcr}}}

\newcommand{\half}{\ensuremath{\frac{1}{2}\,}}

\newcommand{\br}{\ensuremath{{\bf r}}}
\newcommand{\bk}{\ensuremath{{\bf k}}}

\newcommand{\blambda}{\mbox{\boldmath$\lambda$}}
\newcommand{\brho}{\mbox{\boldmath$\rho$}}

\newcommand{\ident}{\ensuremath{\mathbb{I}}}
\newcommand{\kvec}{\ensuremath{\vec{k}}}

\newcommand{\lw}{\blambda_w}
\newcommand{\lv}{\blambda_v}

\newcommand{\rw}{\brho_w}
\newcommand{\rv}{\brho_v}

\newcommand{\evmu}{\vec{\epsilon}_{\mu}}
\newcommand{\evd}{\vec{\epsilon}_{D}}
\newcommand{\lemu}{\lambda_{\epsilon_{\mu}}}
\newcommand{\led}{\lambda_{\epsilon_{D}}}
\newcommand{\slope}{s}
\newcommand{\bomega}{\mbox{\boldmath$\omega$}}

\newcommand{\phibar}{\bar{\varphi}}
\newcommand{\Ys}{Y^{*}}
\newcommand{\Yst}{\tilde{Y}^{*}}
\newcommand{\mus}{\mu^{*}}
\newcommand{\must}{\tilde{\mu}^{*}}
\newcommand{\Ds}{D^{*}}
\newcommand{\Dst}{\tilde{D}^{*}}
\newcommand{\phis}{\varphi^{*}}
\newcommand{\ws}{w^{*}}
\newcommand{\vs}{v^{*}}
\newcommand{\rprior}{\pi_{N}}

\newcommand{\Swwy}{\Sigma_{\hat{w}\,w_{y}}}
\newcommand{\Slw}{\Sigma_{\lambda_{w}}}
\newcommand{\Swy}{\bar{\Sigma}_{w_{y}}}
\newcommand{\Swb}{\bar{\Sigma}_{\hat{w}}}

\begin{document}
\bibliographystyle{apsrev}

\title{Simulations and cosmological inference: \\A statistical model for power spectra means and covariances}

\author{Michael D. Schneider}
\email{schneider@ucdavis.edu}
\affiliation{Department of Physics, University of California, One Shields Avenue, Davis, CA 95616, USA.}

\author{Lloyd Knox}
\affiliation{Department of Physics, University of California, One Shields Avenue, Davis, CA 95616, USA.}

\author{Salman Habib}
\affiliation{T-8, MS B285, Los Alamos National Laboratory, Los Alamos, NM 87545, USA.}

\author{Katrin Heitmann}
\affiliation{ISR-1, MS D466, Los Alamos National Laboratory, Los Alamos, NM 87545, USA.}

\author{David Higdon}
\affiliation{ CCS-6, MS F600, Los Alamos National Laboratory, Los Alamos, NM 87545, USA.}

\author{Charles Nakhleh}
\affiliation{Pulsed Power Sciences, Sandia National Laboratories, Albuquerque, NM 87185, USA.}

\date{\today}

\preprint{LA-UR-08-0730}


\begin{abstract}
We describe an approximate statistical model for the sample variance distribution of the non-linear matter power spectrum that can be calibrated from limited numbers of simulations.  Our model retains the common assumption of a multivariate Normal distribution for the power spectrum band powers, but takes full account of the (parameter dependent) power spectrum covariance.  The model is calibrated using an extension of the framework in~\textcite{habib07} to train Gaussian processes for the power spectrum mean and covariance given a set of simulation runs over a hypercube in parameter space.  We demonstrate the performance of this machinery by estimating the parameters of a power-law model for the power spectrum.  Within this framework, our calibrated sample variance distribution is robust to errors in the estimated covariance and shows rapid convergence of the posterior parameter constraints with the number of training simulations.
\end{abstract}


\maketitle

\section{Introduction}

The indirect nature of most cosmological observations usually requires numerical simulations of the data in order to infer constraints on cosmological models.  For parameter inference from the cosmic microwave background (CMB), galaxy and weak lensing surveys, and the Lyman~$\alpha$ forest, the required simulations can be computationally expensive in order to capture the relevant physics, noise sources, and dynamic range.  The computational demands for future observations will only increase as more accurate theoretical predictions are required to match the reduced errors in the data.  In response to this foreseen bottleneck, several tools have recently been under development to reduce computational costs by emulating the output of cosmological simulations for the CMB and galaxy surveys given a training set of simulations~\cite{pico,pico2,cosmonet,cosmonet2,heitmann06,habib07}.  These tools have been aimed at producing fast estimates of the mean simulation output, but often the error distribution for the data also needs to be inferred from simulations.  

Typically, error models are constructed by running many realizations of a forward simulation of the data (or a compressed version of the data) at a fixed point in the model parameter space.  These multiple realizations can be used, for example, to construct covariance matrix estimates for use in inferring cosmological parameter distributions given the data.  If the error distribution is parameter dependent, then many more forward simulations run with varying input parameters could be required~\cite{seljak03}.

We propose a unified framework for combining estimates of both the data mean and covariance matrix from the same set of simulations for cosmological parameter inference.  Our framework uses an efficient algorithm to interpolate between simulations run at sparse locations in parameter space and allows for propagation of interpolation errors into the inferred cosmological parameter constraints.  This is an extension of the method in Ref.~\cite{habib07} in that sample covariance estimates at several points in parameter space are interpolated along with the sample mean estimates previously considered.  By requiring sample covariance matrices to be computed at many points in parameter space, our model might appear to require a large increase in the computational resources.  However, we also outline a general method to jointly constrain the covariance matrices for different parameters with the combined simulation realizations covering the whole parameter space.  We focus our validation tests on the prerequisite step of demonstrating the statistical framework when the covariances are already known.  

Our model also provides a tool to determine whether the parameter dependence of the errors is important in any given application and a way to incorporate this parameter dependence when it is important (which are issues that can never be addressed by using jacknife covariance estimates from the data).  The parameter dependence of the errors is likely unimportant in any application where the parameters are known {\it a priori} to be tightly constrained.  However, it may not be clear in any given application what constitutes ``tight'' constraints for the purposes of this approximation.  When the parameters are not tightly constrained, we expect it will probably be important to model the parameter dependence of the errors whenever performing inference from a reduced statistic of the data (because residual parameter dependence of the data can be absorbed into the error distribution for the reduced statistic).  We focus on the non-linear matter power spectrum in this paper as an example of this type of situation.  Because the non-linear matter distribution is non-Gaussian, the power spectrum is not a sufficient statistic and the variance of the power spectrum receives contributions from the (parameter dependent) connected four-point function.  It has already been shown~\cite{sefusatti06} that the joint covariance of the two-point and three-point functions of the non-linear galaxy distribution has non-trivial and significant parameter dependence.


A simple example where the parameter-dependence of the errors is important is the measurement of the quadrupole of the CMB power spectrum (which has received considerable interest after WMAP reported a value somewhat lower than expectations). The dominant error on the quadrupole actually depends on the value of the quadrupole itself.  So, a naive analysis where one might attempt to construct the error distribution by running Monte Carlo simulations at a fixed point in parameter space would severely bias the inferred value of the quadrupole.   

In fact, the properties of the large-scale CMB are simple enough that it is easy to analytically solve for the error distribution of the CMB quadrupole~(e.g. Ref.~\cite{bond98}) or, using a sampling approach, even
calculate the multivariate distribution of a whole set of multipole power amplitudes~\cite{wandelt04}.  However, in most situations it is likely that the only recourse is to learn about the error distribution from simulations.  For example, the CMB power spectrum error distribution can no longer be calculated analytically once systematic errors and foreground modelling are included, yet the parameter dependence of the error distribution is likely to remain important.  This will not be the case in general, and the importance of modelling the noise variation over the parameter space will have to be decided on a case-by-case basis.

We explain our framework in the context of performing parameter inference from the non-linear matter power spectrum and have therefore limited the model for the (reduced) data error distribution to a multivariate Normal.  This model could be extended, for example, by considering a mixture of multivariate Normal distributions.  We have otherwise kept a general framework that can be applied to a wide array of applications.  

This paper is organized as follows.  In Section~\ref{sc:DMps} we give some background on the statistical properties of the dark matter power spectrum that serve as motivation for our framework.  In Section~\ref{sc:framework} we describe our model for the power spectrum sample variance distribution and how to calibrate the model using simulations.  We then derive the joint likelihood of the simulation outputs and observed power spectrum for performing parameter estimation.  We test the performance of this framework with a toy model for the power spectrum in Section~\ref{sc:validation}.  In Section~\ref{sc:conclusions}, we summarize our results and outline future directions of this work.  A guide to the notation is given in Appendix~\ref{sc:notation}, a covariance matrix parameterization that fits in our framework is given in Appendix~\ref{sc:covparam}, and details of the likelihood calculation and evaluation are given in Appendices~\ref{sc:emlike}, \ref{sc:proposal}, \ref{sc:priors}, and \ref{sc:covmatexpressions}.

\section{\label{sc:DMps}Dark matter power spectrum}

The primary difficulty in calculating theoretical predictions of the matter distribution (when gas dynamics are neglected) is accounting for non-linear gravitational evolution of the matter density fluctuations.  The only known way to obtain reasonably accurate predictions is by running N-body numerical simulations (although perturbation theory has had some success over a limited range of length scales~\cite{bernardeau02,matsubara07}).  Because two-point functions are ubiquitous in the analysis of galaxy and weak-lensing data, substantial effort has gone into obtaining accurate predictions of the mean of estimators of the dark matter power spectrum~\cite{heitmann05,heitmann07a}.  On the other hand, the error distributions of these power spectrum estimators are much less developed.  Using N-body simulations, \textcite{meiksin99} and \textcite{scoccimarro99} showed that non-linear evolution leads to strong correlations in the band-averaged power spectrum.  \textcite{cooray00} reproduced this result using the halo model and forecast that the non-linear corrections to the power spectrum covariance led to a $\sim15$\% increase in parameter error bars from a fiducial all-sky weak lensing survey.   Using ray-tracing through N-body simulations, \textcite{semboloni06} have shown similar increases to the weak lensing power spectrum variance and correlations due to non-linear evolution.

On a finite or masked region of the sky the window function further modifies the covariance structure of power spectrum estimators.  \textcite{hamilton05} found that the coupling of Fourier modes due to non-linear evolution induces a significant increase in the power spectrum variance when windows are applied to the dark matter density calculated from N-body simulations.  
If ignored, these corrections to the power spectrum covariance could lead to biases and underestimates in inferred cosmological parameter constraints.  Preliminary forecasts have shown that improved modelling of the power spectrum covariance is important in understanding the cosmological information in the non-linear power spectrum~\cite{rimes05b,neyrinck06b}.  Ideally, these effects would be understood by generating mock survey catalogues~\cite{sefusatti06}.  But, this approach quickly becomes computationally prohibitive if we try to run multiple survey simulations for different cosmological models to capture the full parameter dependence of the non-linear dark matter distribution.  We address this problem by extending the methods of Refs.~\cite{habib07}~and~\cite{heitmann06} to build a statistical formulation to accurately model both the power spectrum mean and covariance over parameter space given a fixed number of simulated power spectrum realizations.  We use the scatter between realizations to infer the power spectrum sample variance distribution for a given cosmology, which we then interpolate over the rest of parameter space.  

The non-linear evolution of the dark matter density skews the one-point probability distribution away from its Gaussian initial condition.  As a result, the power spectrum is no longer a sufficient statistic for describing the density field.  An alternative approach to estimating cosmological parameters from the non-linear dark matter distribution could therefore be to model the non-Gaussian one-point distribution directly or to devise alternative summary statistics that capture additional or complementary information to the power spectrum~\cite{taylor01,takada02}.  However, we will not explore this line of inquiry in this paper.

\section{\label{sc:framework}Statistical framework}

We confine our investigation to the distribution of shell-averaged power spectrum estimators of the form,
\begin{equation}\label{eq:psestimator}
	\hat{P}(k_{i}) = \frac{1}{V}\int_{{\rm S}_{i}} \frac{d^{3}k}{V_{{\rm S}_{i}}}\, 
	\delta^{*}(\bk)\delta(\bk),
\end{equation}
where $\delta(\bk)$ is the Fourier transform of the matter density contrast $\delta(\br) = (\rho(\br)-\bar{\rho})/\bar{\rho}$, $V$ is the survey volume, and S$_{i}$ is a spherical shell in $k$-space with radius centered at $k_{i}$.  The shell averaging exploits the assumed isotropy of the density field and reduces the variance of the power spectrum estimator if $\delta(\br)$ is Gaussian.  On large scales, $\delta(\br)$ is indeed expected to be Gaussian and this reduced variance is a prime motivation for constructing power spectrum estimators of the form given in Eqn.~(\ref{eq:psestimator}). 
In the Gaussian case, $\hat{P}$ is a sum of squares of Gaussian variates, and thus Wishart distributed ({\it i.e.} the marginal distributions of each band power are $\chi^{2}$).  The variance of $\hat{P}$ then decreases as one over the number of modes in the shell (as the number of degrees of freedom increase).  However, if $\delta(\br)$ is non-Gaussian in general there will be a non-zero connected 4-point function contributing to the variance of $\hat{P}$, which does not decrease in amplitude with increasing number of modes in the shell~\cite{meiksin99,scoccimarro99}.  The connected 4-point function also introduces correlations in the power spectrum, which are enhanced by the band-averaging (when the Gaussian contribution to the variance is reduced while the off-diagonal covariance remains constant).  	

\subsection{\label{sc:model}Model for the sample variance distribution}
The Central Limit Theorem guarantees that the Normal distribution will be a valid approximation for the distribution of $\hat{P}$ from Eqn.~(\ref{eq:psestimator}) as long as there are a large number of modes in each band power~\cite{szapudi99}.  This approximation will break down on the largest scales of a survey (where only a few modes can be measured), but this could be mitigated by using wider bins.  Alternatively, an exact likelihood could be used if the survey is big enough that the largest scales probe fluctuations in the linear regime.  Therefore, for a given vector of wavenumbers $\kvec=\{ k_1,k_2,\dots,k_{n_{y}}\}$ (where $n_{y}$ is the number of bands), we model the power spectrum sample variance distribution as, 
\begin{equation}\label{eq:mvmodel}
  y(\kvec,\theta) \sim {\rm N}(\mu(\kvec;\theta),\Sigma_{y}(\theta)).
\end{equation}
That is, the observed power spectrum $y$ in bands $\kvec$ for cosmological parameters $\theta$, is assumed to be a random sample from a multivariate-Normal distribution with mean vector $\mu(\kvec;\theta)$ and covariance matrix $\Sigma_{y}(\theta)$ (which has dimensions $n_{y}\times n_{y}$).  We allow for an arbitrary covariance matrix, including the strong correlations and parameter dependence generated by non-linear evolution.  In general it is desirable to reduce the number of components of $\Sigma_{y}(\theta)$ whose $\theta$-dependence must be modelled.  We will denote this subset of components as a column-vector, $D(\kvec;\theta)$, so that $\Sigma_{y}=\Sigma_{y}(D(\kvec;\theta))$.  $D(\kvec;\theta)$ could be, for example, the eigenvalue spectrum with $\theta$-independent eigenvectors assumed for $\Sigma_{y}$.  See Appendix~\ref{sc:covparam} for an explicit example of a paremeterization of the covariance matrix that makes our framework tractable.  

Note that Eqn.~(\ref{eq:mvmodel}) models the distribution of the power spectrum estimator given the parameters as a Gaussian, which does not necessarily imply that the distribution of the true power spectrum given the estimator is Gaussian\footnote{This implication holds only if the parameters are the true band powers and a uniform prior is assumed for the true band powers.}.  In this sense, the model in Eqn.~(\ref{eq:mvmodel}) is quite general.  

\subsection{Calibration from simulations}

We use a fixed number of stochastic simulations of $y(\kvec,\theta)$ at several values of $\theta$ to calibrate the model for the sample variance distribution in Eqn.~(\ref{eq:mvmodel}).  The first step is to choose a set of values of $\theta$ that will cover the region of parameter space we wish to explore while using as small a number of simulation runs as possible.  We refer to this choice as the {\it simulation design}.  Second, we need a way to interpolate the model for the sample variance to new regions of parameter space where no simulations have been run.  We call this the {\it simulation emulator}.

\subsubsection{\label{sc:simdesign}Simulation design}

We follow Section~II.B of Ref.~\cite{habib07} to construct the simulation design as an orthogonal array Latin hypercube sample~\cite{leary03,morris95,tang93,welch85,ye00}.   We begin by specifying a hyper-rectangle in parameter space over which we wish to run simulations.  The parameter axes are then rescaled to give a unit hypercube so that all parameters are subsequently defined on the interval $(0,1)$.  We use the {\sc R} package~\cite{R} {\tt lhs}~\cite{Rlhs} to compute the Latin hypercube sample given the number of design points, $n_{d}$.

For a given $\theta$ we assume a single simulation run gives a random realization of $y(\kvec,\theta)$.  We then run $n_{r_{i}}$ realizations at each design point $i=1,\dots,n_{d}$ for a total of $m\equiv \sum_{i=1}^{n_{d}}n_{r_{i}}$ simulation runs, giving output $Y_{ij}=$~$j$th realization of $y(\kvec,\theta_{i})$ with $\, j=1\,\dots,n_{r_{i}}$.  

In what follows, we use the $*$ superscript to denote simulation outputs for the design settings so that $\left\{Y_{ij}\right\}\equiv Y^{*}$.  We will also find it convenient to label the parameters for the sample variance distribution of $Y^{*}$ at the design points as $\mus$ and $\Ds$ (each of length $n_{y}n_{d}\equiv q$).
Following Ref.~\cite{habib07} and to simplify later prior specifications, we center $\mus$ and $\Ds$ by the constant vectors $\mu_{c}(\kvec)$ and $D_{c}(\kvec)$ to have zero mean and then re-scale each by a single number ($\mu_{c}$ and $D_{c}$) to give unit variance (over the set of simulation runs),
\begin{eqnarray}\label{eq:simdecomposition}
	\tilde{\mu}(\kvec;\theta_{i}^{*}) &\equiv& \left(\mu(\kvec;\theta_{i}^{*})-
	\mu_{c}(\kvec)\right)/\mu_{s}, \nonumber\\
	\ln\left(\tilde{D}(\kvec;\theta_{i}^{*})\right) &\equiv& \left(\ln\left(D(\kvec;\theta_{i}^{*})\right)-
	D_{c}(\kvec)\right)/D_{s},
\end{eqnarray}
where $\theta_{i}^{*}$ denotes the input settings at the $i$th design point ($i=1,\dots,n_{d}$).  We transform to the logarithm of $D$ because our interpolation method requires support over the entire real line (while $D$ has only positive support if $D$ is the eigenvalue spectrum or is as defined in Eqn.~\ref{eq:covparam}).  If a different parameterization of the $\theta$-dependence of $\Sigma_{y}$ gives a non-positive $D$, other mappings of $D$ to the real line can be substituted here.  

If the number of realizations at each design point, $n_{r_{i}}$, is sufficiently large, we can construct a simplified simulation emulator by first reducing the simulation design runs to sample mean and covariance estimates at each design point.  This allows us to reduce the computational complexity of the emulator by inferring the emulator parameters directly from the sample means and covariances.  We use this simplified emulator for the examples in Section~\ref{sc:validation} with the added assumption that the sample means and covariances are perfect estimates of the true means and covariances.  The number of realizations at each design point, $n_{r}$, required to make this approximation valid for the covariance can be many times the number of power spectrum bands, $n_{y}$.  More optimized techniques for estimating the power spectrum covariance from simulations might also be helpful in some applications~\cite{pope07}.


\subsubsection{Simulation emulator}

We can further reduce the number of components to model by performing a principal component (PC) analysis on the scaled means, $\tilde{\mu}(\kvec,\theta^*)$, and variances, $\tilde{D}(\kvec,\theta^{*})$, of the design simulations.  Following Ref.~\cite{habib07}, we perform a singular value decomposition on the $n_{y}\times n_{d}$ matrix of simulation sample means at each design setting, $\left[\tilde{\mu}^*\right]=\mathsf{UBV}^T$ where $\mathsf{U}$ has dimension $n_{y}\times p$ ($p\equiv\text{min}(n_{y},n_{d})$) with $\mathsf{U}^{T}\mathsf{U}=\ident_{p}$, $\mathsf{V}$ has dimension $n_{d}\times p$ with $\mathsf{V}^{T}\mathsf{V}=\ident_{p}$, $\mathsf{V}\mathsf{V}^{T}=\ident_{n_{d}}$, and $\mathsf{B}$ ($p\times p$) is a diagonal matrix of singular values.
We then decompose $\tilde{\mu}^{*}$ in the basis vectors, $\Phi_{\mu}=\mathsf{U}$ and weights $w=\mathsf{BV}^{T}$ so that $\Phi_{\mu}^{T}\Phi_{\mu}=\ident_{p}$ (with an analogous decomposition for $[\ln\tilde{D}^*]$)\footnote{Ref.~\cite{habib07} use the alternate weighting $\Phi_{\mu}=\frac{1}{\sqrt{n_{d}}}\mathsf{U}\mathsf{B}$ and $w=\sqrt{n_{d}}\mathsf{V}^{T}$ so that $\frac{1}{n_{d}}w^{T}w=\ident_{n_{d}}$}.  Retaining only the first $p_{\mu}$ and $p_{D}$ columns of $\Phi_{\mu}$ and $\Phi_{D}$,
\begin{eqnarray}\label{eq:pcdecomp}
	\tilde{\mu}(\kvec;\theta) &=& \sum_{i=1}^{p_{\mu}} \Phi_{\mu,i}(\kvec)\,w_{i}(\theta) + \evmu, \nonumber\\
	\ln\left(\tilde{D}(\kvec;\theta)\right) &=& \sum_{i=1}^{p_{D}} \Phi_{D,i}(\kvec)\,v_{i}(\theta) + \evd,
\end{eqnarray}
where $p_{\mu},p_D \le n_{y}$, $\Phi_i$ is the $i$th column of $\Phi$, $w_{i}$ and $v_{i}$ are (parameter dependent) basis weights, and $\evmu,\evd$ are independent and identically distributed (i.i.d.)~Normal variates parameterizing the error in the truncation of the principal component (PC) decomposition.  

The parameter dependence of the likelihood has now been isolated into a set of $p_{\mu}+p_{D}$ basis weights for the power spectrum mean and ``log-variance''.  To find a model that fits all the simulation design runs, we 
again follow Ref.~\cite{habib07} and model the basis weights as Gaussian processes (GP) over the prior parameter space,
\begin{eqnarray}\label{eq:gpdists}
	w_{i}(\theta) &\sim& {\rm GP}\left(0,\Sigma_{w}(\theta;\lambda_{w,i},\brho_{w,i})\right)
	\qquad i=1,\dots,p_{\mu},\nonumber\\
	v_{i}(\theta) &\sim& {\rm GP}\left(0,\Sigma_{v}(\theta;\lambda_{v,i},\brho_{v,i})\right)
	\qquad i=1,\dots,p_{D},
\end{eqnarray}
where,
\begin{equation}\label{eq:GPcov}
  \Sigma_{X}(\theta,\theta';\lambda_{X,i},\brho_{X,i}) = \lambda_{X,i}^{-1}\prod_{\ell=1}^{p_{\theta}} \rho_{X,i\ell}^{4(\theta_{\ell}-\theta_{\ell}')^{2}}
\end{equation}
gives the covariance of the GP for weight $i$ between parameter values $\theta$ and $\theta'$ with precision $\lambda_{X,i}$ and correlations (over the parameter space) $\brho_{X,i}$.

From Eqns.~(\ref{eq:pcdecomp}) and (\ref{eq:gpdists}) we can now derive the sampling models for the parameters $\must$ and $\Dst$.  Let $\mu^{*}$ and $D^{*}$ denote the $n_{y}n_{d}\equiv q$ column vectors obtained by concatenating the sample means and variances at each design point.  Further, let $w^{*}$ and $v^{*}$ denote the PC weights for $\must$ and $\ln(\Dst)$ evaluated at the design points.  Then, from the i.i.d. Normal model for $\lemu$ and $\led$,
\begin{eqnarray}\label{eq:GPmodeConditionals}
  \must|w^*,\lemu &\sim& {\rm
    N}(\Phi_{\mu}w^*,\lemu^{-1} \ident_{q}),\nonumber\\
  \ln\left(\Dst\right)|v^*,\led &\sim& {\rm
    N}(\Phi_{D}v^*,\led^{-1}\ident_{q}).
\end{eqnarray}
Restricted to the design points, the GP models give Normal priors for $w^{*}$, $v^{*}$, and $\phis$,
\begin{eqnarray}\label{eq:despriors}
	w^{*} &\sim& {\rm N}\left(0,\Sigma_{w}^{*}(\blambda_{w},\brho_{w})\right), \nonumber\\
	v^{*} &\sim& {\rm N}\left(0,\Sigma_{v}^{*}(\blambda_{v},\brho_{v})\right),
\end{eqnarray}
where $\Sigma_{w}^{*}$, $\Sigma_{v}^{*}$, and $\Sigma_{\varphi}^{*}$ are the extension of Eqn.~(\ref{eq:GPcov}) to the design points for each mode amplitude (see Appendix~\ref{sc:covmatexpressions}).  

The marginal distribution for $\must$ is,
\begin{eqnarray}\label{eq:wstarmarg}
	&&\pi(\must|\lemu,\blambda_{w},\brho_{w}) = \nonumber\\
	&&\quad\int dw^{*}\, 
	\pi(\must|\lemu,w^{*})\cdot \pi(w^{*}|\blambda_{w},\brho_{w}).
\end{eqnarray}
We use the intermediate result from~Eqn.~(21) of Ref.~\cite{habib07} along with the definition $\hat{w}\equiv\Phi_{\mu}^{T}\must$ to get, 
\begin{eqnarray}
	&&\pi(\must|\lemu,w^{*}) \propto \lemu^{\frac{n_{d}p_{\mu}}{2}}
	\exp\left[-\half\lemu(w^{*}-\hat{w})^{T}(w^{*}-\hat{w})\right]
	\nonumber\\
	&&\times \lemu^{n_{d}(n_{y}-p_{\mu})/2}\exp\left[
	-\half\lemu \tilde{\mu}^{*T}(\ident_{q}-\Phi_{\mu}\Phi_{\mu}^{T})\must\right],
	\label{eq:reddesign}\\
	&&\equiv \pi(w^{*}|\hat{w},\lemu)\cdot
	\rprior(\must|\lemu),\nonumber
\end{eqnarray}
with an analogous result for $\pi(\ln(\Dst)|\led,v^{*})$.  It is now straightforward to perform the integral in Eqn.~(\ref{eq:wstarmarg}),
\begin{eqnarray}
	\pi(\must|\lemu,\blambda_{w},\brho_{w}) &=& 
	\pi(\hat{w}|\lemu,\blambda_{w},\brho_{w})\nonumber\\
	&\times& \rprior(\must|\lemu),
\end{eqnarray}
where
\begin{eqnarray}
	\hat{w}|\lemu,\lw,\rw &\sim& {\rm N}\left(0,\lemu^{-1}\ident + 
	\Sigma_{w}^{*}(\lw,\rw)\right).
\end{eqnarray}
Similarly for $\ln(\Dst)$,
\begin{eqnarray}
	\pi(\ln(\Dst)|\led,\blambda_{v},\brho_{v}) &=& 
	\pi(\hat{v}|\led,\blambda_{v},\brho_{v})\nonumber\\
	&\times& \rprior(\ln(\Dst)|\led),
\end{eqnarray}
with
\begin{eqnarray}
	\hat{v}|\led,\lv,\rv &\sim& {\rm N}\left(0,\led^{-1}\ident + 
	\Sigma_{v}^{*}(\lv,\rv)\right),
\end{eqnarray}
and $\hat{v}\equiv \Phi_{D}^{T}\ln(\Dst)$.

We calibrate the emulator by using Markov Chain Monte Carlo (MCMC) to draw samples from the posterior of the GP model parameters given the design runs, $\pi(\bomega | \Ys)$ ($\bomega\equiv\left\{\lemu,\lw,\rw,\led,\lv,\rv\right\}$).  For the ``simplified emulator'' described at the end of Section~\ref{sc:simdesign}, this posterior factors so the parameters for the power spectrum mean and variance can be calibrated separately,
\begin{eqnarray}
  &&\pi(\bomega|\tilde{\mu}^*,\tilde{D}^*) \propto
  \label{eq:simpgplike}\\
  &&\left[\pi(\must|\lemu,\lw,\rw)\cdot
  \pi(\lemu) \cdot \pi(\lw)\cdot \pi(\rw)\right]
  \nonumber\\
  &\times&\left[\pi(\ln(\Dst)|\led,\lv,\rv)\cdot
  \pi(\led) \cdot \pi(\lv)\cdot \pi(\rv)\right].\nonumber
\end{eqnarray}
By sampling from this posterior, we can propagate the error in the calibration of the models for $\mu$ and $D$ from our limited set of simulation runs.  For the simplified emulator likelihood in Eqn.~(\ref{eq:simpgplike}), the model for the mean is identical to that in Ref.~\cite{habib07}.  Explicit expressions for the full likelihood and priors are given in Appendices~\ref{sc:emlike},~\ref{sc:priors},~and~\ref{sc:covmatexpressions}.

\subsection{Cosmological parameter estimation}

We now consider how to use our simulation-calibrated model for the sample variance distribution to estimate cosmological parameters from an observation of the power spectrum, denoted $y(\kvec)$.  For complete error propagation, our goal is to compute the joint posterior
$\pi\left(\theta_{0},\bomega|y,Y^{*}\right)$, or, if using the ``simplified emulator,'' $\pi\left(\theta_{0},\bomega|y,\mu^{*},D^{*}\right)$, where $\theta_{0}$ are the ``true'' parameters that generated the observation $y(\kvec)$.  

First, we decompose the mean and variance of the data error distribution into the same bases as the design runs.  The model for the sample variance distribution in Eqn.~(\ref{eq:mvmodel}) becomes
\begin{equation}
	\tilde{y}|w(\theta_{0}),v(\theta_{0}) \sim {\rm N}\left(\Phi_{\mu}w(\theta_{0}),
	W_{y}^{-1}(v(\theta_{0}))\right),
\end{equation}
where $W_{y}^{-1}(v(\theta_{0}))\equiv \Sigma_{y}\left(\exp\left(D_{s}\Phi_{D}v(\theta_{0})+D_{c}\right)\right)/\mu_{s}^{2}$.  Note that we model the mean and variance of the observations as perfectly described by the PC weights $w(\theta_{0})$ and $v(\theta_{0})$, without the error terms that were included in the decomposition of the simulation means and variances in Eqn.~(\ref{eq:pcdecomp}).  Next, to simplify the expression for marginalizing over $w$, we rewrite this distribution in terms of 
\[
\hat{w}_y(\theta) \equiv \left(\Phi_{\mu}^TW_{y}(\theta)\Phi_{\mu}\right)^{-1} \Phi_{\mu}^T W_{y}(\theta) y
\]
in analogy with Eqn.~(\ref{eq:reddesign}).  However, because $W_{y}$ depends on $\theta_{0}$, we must be careful to preserve all the normalization factors.  The exact relation is:
\begin{widetext}
\begin{eqnarray}
	L(y|w,v) &=& \left[(2\pi)^{n_{y}}\left|W_{y}^{-1}\right|\right]^{-1/2} \,
	\exp\left\{-\half\left(\tilde{y}-\Phi_{\mu}w\right)^{T}W_{y}\left(\tilde{y}-\Phi_{\mu}w\right)\right\} 
	\nonumber\\
	&=& \left[(2\pi)^{p_{\mu}}\left|\Phi_{\mu}^{T}W_{y}\Phi_{\mu}\right|^{-1}\right]^{-1/2}\,
	\exp\left\{-\half\left(w-\hat{w}_{y}\right)^{T}\Phi_{\mu}^{T}W_{y}\Phi_{\mu}
	\left(w-\hat{w}_{y}\right)\right\}\nonumber\\*
	&\times& (2\pi)^{-(n_{y}-p_{\mu})/2}\left|W_{y}\right|^{1/2}
	\left|\Phi_{\mu}^{T}W_{y}\Phi_{\mu}\right|^{1/2}\,
	\exp\left\{-\half\left(\tilde{y}-\Phi_{\mu}\hat{w}_{y}\right)^{T}W_{y}
	\left(\tilde{y}-\Phi_{\mu}\hat{w}_{y}\right)\right\}\nonumber\\
	&\equiv& L(\hat{w}_{y}|w,v)\cdot \rprior(y|v).
\end{eqnarray}
\end{widetext}
The first line of the final result is a properly normalized Gaussian distribution in $w$, while the second line is independent of $w$.  The priors on the PC weights for the data are,
\begin{eqnarray}
	w(\theta_{0}) &\sim& {\rm N}\left(0,\Sigma_{\lambda_{w}}\right)\,\text{and}\nonumber\\
	v(\theta_{0}) &\sim& {\rm N}\left(0,\Sigma_{\lambda_{v}}\right),
\end{eqnarray}
where,
\begin{eqnarray}
	\Sigma_{\lambda_{w}} &=& {\rm diag}\left(\lambda_{w_{i}}^{-1}\right)\qquad
	(p_{\mu}\times p_{\mu})\,\text{and} \nonumber\\
	\Sigma_{\lambda_{v}} &=& {\rm diag}\left(\lambda_{v_{i}}^{-1}\right)\qquad
	(p_{D}\times p_{D})\nonumber.
\end{eqnarray} 

\begin{widetext}
The joint likelihood for the data and simulation outputs can be constructed by multiplying the individual likelihoods and marginalizaing over the variables for the mean and covariance (weighted by their prior distributions),
\begin{eqnarray}
	L(y,\Ys|\theta_{0},\bomega) &=&\int\int d\mus\,d\Ds\,
	L(\Ys|\mus,\Ds)\,\int\int dw^{*}\,dw_{0}\,\int\int dv^{*}\,dv_{0}\, 
	 L(y|w_{0},v_{0})\nonumber\\
	&\times&\pi(\mus|\ws,\lemu)\cdot \pi(\Ds|\vs,\led)
	\cdot\pi(w^{*},w_{0}|\theta_{0},\lw,\rw)
	\cdot\pi(v^{*},v_{0}|\theta_{0},\lv,\rv).\nonumber
\end{eqnarray}
The integrals over $w^{*},w_{0}$ and $\vs$ can be performed analytically, giving,  
\begin{eqnarray}\label{eq:finallike}
	L(y,\Ys|\theta_{0},\bomega) &=&\int\int d\mus\,d\Ds\,
	\int dv_{0}\,L(\Ys|\mus,\Ds)\\
	&\times&\, \pi(\hat{w}_{y},\hat{w}|v_{0},\theta_{0},\bomega)\cdot 
	\pi_{N}(\mus|\lemu)\cdot \pi_{N}(y|v_{0})\cdot
	\pi(v_{0},\hat{v}|\theta_{0},\bomega) \cdot 
	\pi_{N}(\Ds|\led),\nonumber
\end{eqnarray}	
where,
\begin{equation}\label{eq:wjointlike}
  \left(
    \begin{array}{c}
      \hat{w} \\
      \hat{w}_y
    \end{array} \right)
  \sim {\rm N}\left( \left(\begin{array}{c} 0 \\ 0 \end{array}\right), 
    \left[ \left(\begin{array}{cc}
        \lemu^{-1}\ident_{n_dp_{\mu}} & 0 \\
        0 &
        (\Phi_{\mu}^TW_{y}\Phi_{\mu})^{-1} \end{array}\right)
    + \left(\begin{array}{cc}
        \Sigma_{\hat{w}} & \Sigma_{\hat{w}\,w_y} \\
        \Sigma_{\hat{w}\,w_y}^T & \Sigma_{\lambda_{w}}
      \end{array}\right)\right]\right),
\end{equation}
\begin{equation}\label{eq:vjointlike}
  \left(
    \begin{array}{c}
      \hat{v} \\
      v(\theta_{0})
    \end{array} \right)
  \sim {\rm N}\left( \left(\begin{array}{c} 0 \\ 0 \end{array}\right), 
    \left[ \left(\begin{array}{cc}
        \led^{-1}\ident_{n_dp_{D}} & 0 \\
        0 & 0 \end{array}\right)
    + \left(\begin{array}{cc}
        \Sigma_{\hat{v}}  & \Sigma_{\hat{v}\,v} \\
        \Sigma_{\hat{v}\,v}^T & \Sigma_{\lambda_{v}}
      \end{array}\right)\right]\right).
\end{equation}  
Eqn.~(\ref{eq:finallike}) is simplified further in Appendix~\ref{sc:emlike} and explicit expressions for the covariance matrices are given in Appendix~\ref{sc:covmatexpressions}.

For the simplified emulator (that is conditioned directly on the sample means and variances from the design runs), the integrals over $\mus$ and $D^{*}$ in Eqn.~(\ref{eq:finallike}) can be dropped.  The joint likelihood for the data and the simulation runs in this case is,
\begin{eqnarray}\label{eq:simpemlike}
  L(y,\mus,D^{*}|\theta_0,\bomega)
  = &\int& d^{p_D}v_{0}\,
	\pi(\hat{w}_{y},\hat{w}|v_{0},\theta_{0},\bomega)\cdot 
	\pi_{N}(\mus|\lemu)\cdot \pi_{N}(y|v_{0}) \nonumber\\
	&\times&\pi(v_{0},\hat{v}|\theta_{0},\bomega) \cdot \pi_{N}(\Ds|\led).
\end{eqnarray}
We use this likelihood distribution in an MCMC algorithm to simultaneously constrain the $\theta_{0}$ and the GP parameters.  The details of the likelihood evaluation, the prior distributions on the parameters, and the proposal distributions for our Metropolis-Hastings updates are given in the Appendices.

\end{widetext}

\section{\label{sc:validation}Validation tests}
In this Section we use a toy power-law model for the power spectrum to test the performance of our statistical framework.  We work with a toy model both to speed the computation time involved and to separate issues with the GP calibration from issues with modelling more complicated power spectra and their covariance structures.  Our statistical framework is kept completely general, however, so more sophisticated simulations can be added without further modification.

\subsection{\label{sc:toymodel}Power-law power spectrum model}
We use a two-parameter model for the power spectrum,
\begin{equation}\label{eq:toymodel}
  P(k_i) = A\, k_i^{-\slope},
\end{equation}
characterized by the amplitude, $A$, and slope, $\slope$.  To give sufficient information to distinguish constraints on $A$ and $\slope$, we use $n_{y}=32$ bands in $k$ with $k_{1}=2\pi/450$ and $\Delta k=8k_{1}$.  We set ``true'' values of $A=200$ and $\slope=0.5$, which roughly match the amplitude and shape of the matter power spectrum inside a 450~Mpc/$h$ cubic volume.  

To match our model for the power spectrum distribution, we assume $\hat{P}(\vec{k})$ is multivariate Normal distributed with covariance 
\begin{equation}\label{eq:toycov}
  C = {\rm diag}\left( \frac{2P^2(\vec{k})}{4\pi}\right).
\end{equation}
This is the standard prediction for the covariance of power of a Gaussian random field with $\sim 4\pi$ modes contributing to the power estimate in each $k$-band.  In practice, the number of modes available in each $k$-band increases with the volume of the shell in $k$-space.  However, we assume the same number of modes are used in each band as a way to increase the variance for later validation purposes.  In this model, the decomposition of the covariance as described in Section~\ref{sc:model} is trivial and we set $D(\vec{k};\theta)={\rm diag}(C)$. 

Our ``simulations'' for this model simply involve computing the true power spectrum mean and covariance from Eqns.~(\ref{eq:toymodel}) and (\ref{eq:toycov}).  We neglect the error in sample mean and covariance estimates unless explicitly stated.  In the principal component decompositions, we retain 7 modes in the mean and 2 modes in the log-variance.  We found that our method is numerically stable to retaining modes with small weights (i.e. more modes than necessary in the decomposition), although the MCMC sampling of these weights can be inefficient.  The GP model will automatically determine which weights are active in which directions of parameter space (see Fig.~\ref{fg:rhobox}).  We just have to make sure to use enough modes in the basis decomposition so that we do not lose important features in the response.

\subsection{Results}

\begin{figure}[hb]
	\centerline{
		\scalebox{0.6}{\includegraphics{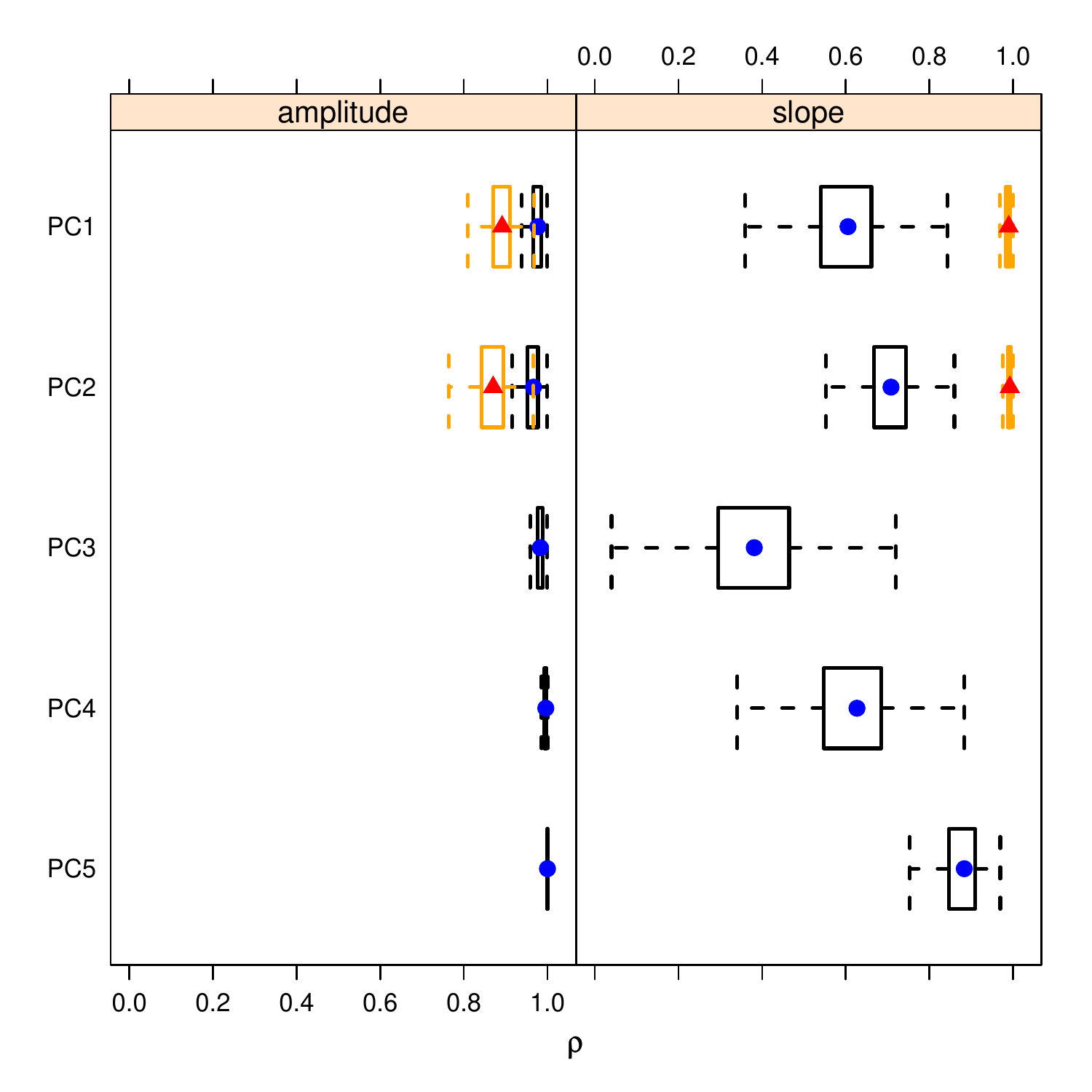}}
	}	
	\caption{\label{fg:rhobox}Boxplots of marginal posterior realizations of the GP correlation parameters for the PC weights of the mean (blue, circles) and covariance (red, triangles) of the power spectrum.  The points indicate the medians of the marginal posterior realizations while the boxes extend from the 1st to the 3rd quartiles.  The bars (frequently called ``whiskers'') indicate the extent of the tails of the distribution and extend to the most extreme sample point that is no more than 1.5 times the box length away from the box.}
\end{figure}

The marginal posterior distributions for the GP correlation parameters $\brho_{w}$ and $\brho_{v}$ are summarized in Fig.~\ref{fg:rhobox}.  The boxes are centered on the medians and extend to the first and third quartiles, while the bars indicate the extent of samples in the tails of the distribution.  A $\rho=1$ indicates a linear interpolation of the surface (perfect correlation) in the given direction of parameter space for the given PC weight, while a $\rho=0$ indicates a rapidly varying surface.  From Eqns.~(\ref{eq:pcdecomp}), (\ref{eq:toymodel}), and (\ref{eq:toycov}), we can see that the parameter dependence of the PC weights is,
\begin{eqnarray}
	w_{i} &=& \phi_{\mu,i}^{T}\,\mu(\kvec;\theta) \sim A\,\left(\phi_{\mu,i}^{T}\kvec^{-\slope}\right),\\
	v_{i} &=& \phi_{D,i}^{T}\,\ln(D(\kvec;\theta) )\sim 
	\ln(A)\,\left(\phi_{D,i}^{T}\ident\right) - 
	 2\slope\,\left(\phi_{D,i}^{T}\ln\kvec\right).\nonumber
\end{eqnarray}
So $w_{i}(\theta)$ is linear in $A$ for fixed $\slope$ and $v_{i}(\theta)$ is linear in $\slope$ for fixed $A$.  This dependence is accurately reflected in the posteriors in Fig.~\ref{fg:rhobox} where $\rho_{w_{i},A}$ and $\rho_{v_{i},\slope}$ are tightly distributed near 1 for all the modes.  Although we retained 7 modes in the decomposition of the mean, $\mu(\kvec)$, only the 5 modes plotted in Fig.~\ref{fg:rhobox} showed active posterior distributions.

\begin{figure}[ht]
	\centerline{
		\scalebox{0.4}{\includegraphics{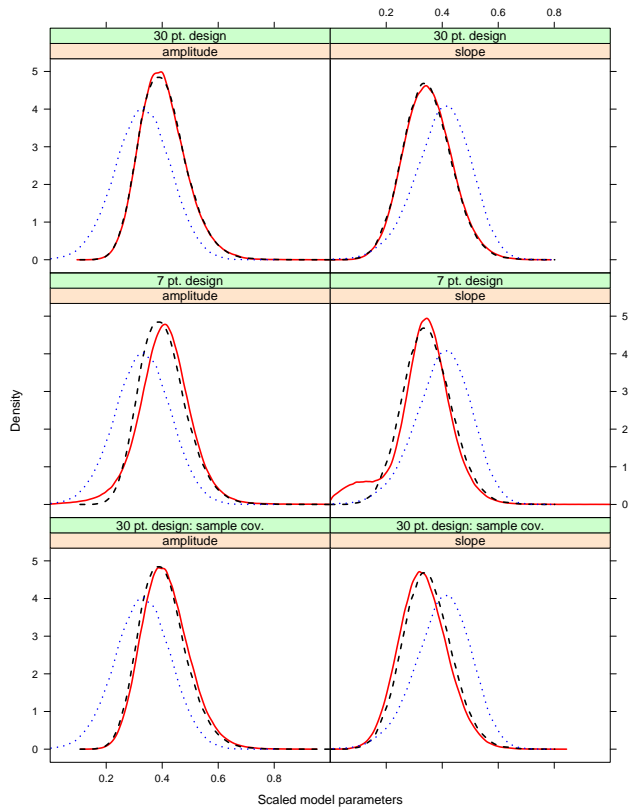}}
	}
	\caption{\label{fg:paramposts}Marginal posteriors of the ``cosmological parameters.''  Black (dashed) is the exact result while red (solid) is the result from our model.  Top: 30 point design.  Middle: 7 point design.  The blue (dotted) lines show the posteriors obtained neglecting the parameter dependence of the covariance.  Bottom: 30 point design using the sample covariance estimated from $n_{r}=32$ realizations at each design point.}
\end{figure}

In Fig.~\ref{fg:paramposts} we show comparison of the marginal parameter posteriors computed using the calibrated power spectrum distribution with the exact result (computed using standard Metropolis-Hastings MCMC).  The top panels show the results for a 30-point simulation design while the middle panels show the same results for a 7-point design.  The 30-point design results are nearly indistinguishable from the exact result, indicating the design points have sufficiently sampled the variation in the mean and covariance response surfaces.  The 7-point design results, however, show noticeable deviations from the exact result.  

\begin{figure}[ht]
	\centerline{
		\scalebox{0.35}{\includegraphics{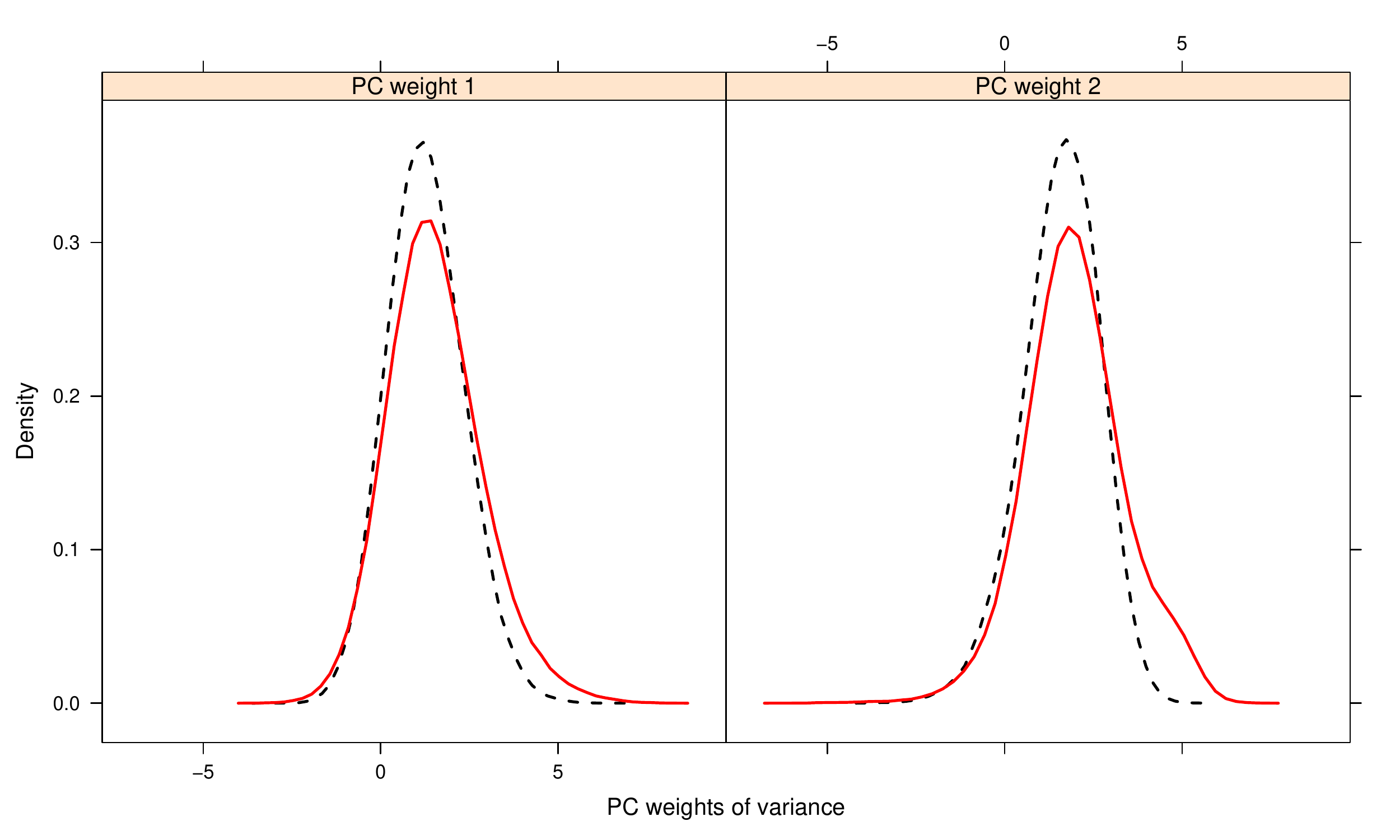}}
	}
	\caption{\label{fg:vipost} Marginal posteriors of the principal component weights for the power spectrum ``variance'' at the true cosmology, $v(\theta_{0})$.  The black (dashed) curves show the posteriors for the 30-point design, while the red (solid) curves show the 7-point design posteriors.}
\end{figure}

The dotted blue lines in the middle panels show the posteriors obtained by fixing the parameters in the covariance to the ``true'' values (so the parameter dependence of the covariance is neglected).   We can see that the 7-point design posteriors are much closer to the exact result than to the fixed-covariance result.  We interpret this as indicating that the parameter dependence of the covariance is still captured, but with more noise than in the 30-point design. The ``bump'' in the tail of the marginal posterior for $\slope$ in the middle panel of Fig.~\ref{fg:paramposts} is an artifact of the interpolation error in this sparse design.  The ``bump'' occurs in a region of parameter space where the GP models attempt to extrapolate from the nearest design point to the edge of our parameter prior region.  However, the 7 points in the design only loosely constrain the GP parameters so the extrapolation is not well-defined.  We have confirmed that a different 7-point design realization can remove the ``bump'' in the $s$ posterior, but only at the expense of larger errors elsewhere in the joint posterior.    Figure~\ref{fg:vipost} shows the marginal posteriors for the variance PC weights for the 30-point and 7-point designs.  This gives a clear illustration of how the posterior distributions broaden (although asymmetrically) as the number of design points is reduced.

The bottom panels of Fig.~\ref{fg:paramposts} show the marginal parameter posteriors when a noisy estimate of the sample covariance is used in the design instead of the perfectly known population covariance.  We used $n_{r}=32$ realizations to estimate the variance at each design point.
While deviations from the exact posteriors can be seen, the match with the exact result is quite close compared to the width of the posterior distributions.

\subsection{Challenges for practical implementation}
Several complexities may arise in applying our method to the analysis of actual galaxy or weak lensing surveys.  A significant challenge for the simplified emulator demonstrated here will likely be the computation of converged covariance matrix estimates at each simulation design point.  
However, the only costs incurred with more design points or band-powers in the observed power spectrum are the increased time for computing the Cholesky factorizations of the covariance matrices in the likelihood (see Appendix~\ref{sc:covmatexpressions}).

For estimating the covariance of the 3-dimensional matter power spectrum from N-body simulations, it was found in Ref.~\cite{meiksin99} that several hundred simulation realizations were needed to obtain converged estimates of the covariance for 20 bands in wavenumber.  If the 128-point design used in Ref.~\cite{habib07} for computing the mean power spectrum is also sufficient sampling for the covariance, then our simplified emulator could possibly require as many as $\sim 128\times 200 = 25600$ runs of an N-body code to calibrate the sample variance distribution of the 3-D dark matter power s	pectrum.  However, we expect a sparser sampling of the covariance would suffice in several directions of the 5-dimensional parameter space used in Ref.~\cite{habib07}.  In addition, once the parameterization of (the $N\times N$) $\Sigma_{y}$ is chosen, our formulation is only concerned with modelling a few of the $N(N+1)/2$ degrees of freedom in the covariance.  It may be possible to estimate the degrees of freedom of interest with substantially fewer power spectrum realizations than are needed to determine the entire covariance matrix.  And, combined with the smoothness assumptions in the GP models, the degrees of freedom in the covariance matrix might be jointly constrained across the simulation design with many fewer realizations than are needed to constrain the covariance at just one point in parameter space.  Finally, because estimates of the mean power spectrum also require several simulation realizations, the parameterized covariance could possibly be constrained without any additional simulation runs.  The techniques proposed in Refs.~\cite{hamilton05}~and~\cite{pope07} for estimating the power spectrum covariance with limited numbers of the N-body simulations could also potentially be useful for our framework.  However, more work may need to be done to accurately capture the effects of the survey window with these methods.  

As detailed in Eqns.~(\ref{eq:finallike}) and (\ref{eq:finallike2}), these difficulties with estimating covariance matrices from simulations may be avoided by conditioning the emulator on the individual power spectrum realizations at each design point.  The potential challenge in this case is performing the Monte Carlo integral over the $(n_{y}\,\sum_{i}n_{r_{i}})$ $\Ds$ components in Eqn.~(\ref{eq:finallike2}).  This is not necessarily a computational obstacle if an appropriate proposal distribution for the Metropolis MCMC algorithm can be found.  Note that according to Eqn.~(\ref{eq:pcdecomp}), $(n_{y}-p_{D})$ components of $\Ds$ at each design point are i.i.d. Gaussian random variates; which should be easy to sample in an MCMC.  That leaves only $(p_{D}\sum_{i}n_{r_{i}})$ correlated components of $\Ds$ to sample.  We found that the prior on $v(\theta)$ is an excellent proposal distribution for computing the integral in Eqn.~(\ref{eq:vjointlike}), and this form may scale easily to more dimensions.

Our toy model in Section~\ref{sc:toymodel} avoided the potentially complicated issue of parameterizing the cosmological parameter dependence of the power spectrum covariance.  While it is straightforward to calculate an eigenvalue spectrum, more general parameterizations will likely be needed for practical application of our method.  
There is a large literature on parameterizing covariance matrices~\cite{manly87,boik02,daniels06,pourahmadi07} that can be applied to this problem, but the choice of parameterization may be a significant complication beyond the toy model studied here.  
We describe how the parameterization of \cite{pourahmadi07} can fit into our framework in Appendix~\ref{sc:covparam}, but this remains untested in a numerical example.  
Because our statistical formulation is insensitive to the choice of parameterization, the only other practical difficulty might come from increased computation time in repeatedly constructing and deconstructing $\Sigma_{y}(D(\kvec;\theta))$.  This will have to be addressed on a case-by-case basis.

\section{\label{sc:conclusions}Conclusions}

We have demonstrated an extension to the statistical model of Ref.~\cite{habib07} to estimate cosmological parameters from the power spectrum using a sample variance distribution calibrated from simulations.  This framework allows modelling of arbitrary, parameter-dependent power spectrum covariance matrices given 
several realizations of the power spectrum at a fixed number of points in parameter space.  We have focused on modelling the covariance of a multivariate Normal model for the estimated power spectrum in order to capture the correlations induced by filtering a Gaussian CMB or galaxy map or from non-linear graviational evolution in the matter power spectrum.  

We tested the calibration of our model from simulations using a toy power-law model for the power spectrum.  In order to focus our tests, we used a simplified emulator that is conditioned on sample means and variances of the simulated power spectra rather than on the individual power spectrum realizations.  For this model, our calibration procedure converges quickly and is quite robust to reducing the number of simulation design points.  We expect that the requirement of computing converged sample covariance estimates at each design point is likely to be a strain on the simulation resources of actual galaxy and weak lensing survey analyses. Therefore, we have described a general formulation of the emulator that allows for constraining parameterized covariance matrices jointly with the other emulator parameters.  Again for our toy model, we have shown that while noisy covariance estimates bias the parameter constraints, the shift is small compared to the width of the parameter posterior distributions.

Our final goal with this work is to develop practical tools to aid in the estimation of cosmological parameters from future measurements of galaxy and cosmic shear power spectra.  As a next step we plan to demonstrate our calibration algorithm using N-body simulations of the dark matter density.  With N-body simulations, our framework provides the means to understand in which regimes modelling of non-linear evolution is important for estimating parameters and, as a related question, how much cosmological information can be extracted from non-linear scales in the dark matter distribution~\cite{taylor01,rimes05b,neyrinck06b}.  The non-trivial effects of the survey window on the power spectrum covariance discussed in Refs.~\cite{hamilton05} and~\cite{sefusatti06} could potentially lead to biases in inferred parameter constraints without the careful modelling allowed by our framework.  In particular, the scaling of the ``beat-coupling'' effect described in Ref.~\cite{hamilton05} with the fundamental modes in a survey implies extra parameter-dependence in the small-scale power spectrum covariance that could be significant in estimating cosmological parameters.  An emulator for N-body simulations will also provide valuable tests of the full emulator formulation presented here that conditions the GP models on the scatter between power spectrum realizations directly.  In this formulation (and a parameterization as in Appendix~\ref{sc:covparam}), it may be possible to model the parameter dependent power spectrum covariance without any more simulations than are needed to accurately estimate the mean power.

\begin{widetext}
\appendix
\section{\label{sc:notation} Notation}
See Table~\ref{tb:notationkey} for the key to the notation used in the paper.
 \begin{table}[hb]
   \begin{center}
     \begin{tabular}{clc}
       \hline \hline 
       Symbol & Description & Value\\
       \hline
       $n_y$  & number of band powers in $k$ & 32 \\
       $p_{\theta}$ & dimensionality of the parameter space & 2 \\
       $n_r$  & number of simulations runs at each design point  & NA\\
       $n_d$  & number of design points & 30,7\\
       $p_{\mu}$ & number of modes in decomposition of $\mu(\kvec;\theta)$ & 7\\
       $p_{D}$ & number of modes in decomposition of $\log(D(\kvec;\theta))$ & 2\\
       $\theta$ & cosmological parameters &\\
       $y(\kvec)$ & observed power spectrum &\\
       $\lemu$ & precision for the error in the PC decomposition of the mean\\
       $\led$ & precision for the error in the PC decomposition of the covariance\\
       $\lw=\left\{\lambda_{w,1},\dots,\lambda_{w,p_{\mu}}\right\}$ & precision of the GP models for the power spectra means &\\
       $\lv=\left\{\lambda_{v,1},\dots,\lambda_{v,p_{D}}\right\}$ & precision of the GP models for the power spectra variances &\\       
       $\rw=\left\{\rho_{w,1},\dots,\rho_{w,p_{\mu}p_{\theta}}\right\}$ & correlations of the GP models for the power spectra means &\\       
       $\rv=\left\{\rho_{v,1},\dots,\rho_{v,p_{D}p_{\theta}}\right\}$ & correlations of the GP models for the power spectra variances &\\              
       \hline \hline
     \end{tabular}
     \caption{\label{tb:notationkey}Key to the notation used in the paper.  The ``Value'' column indicates the values assigned in the validation tests of Section~\ref{sc:validation}.}
   \end{center}
 \end{table}

\section{\label{sc:covparam}Covariance matrix parameterization}

We require a covariance matrix parameterization that is general enough to be applied to a wide array of applications while remaining computationally tractable within our framework.  We focus on the generalized Cholesky decomposition described in Ref.~\cite{pourahmadi07}, although other choices may certainly be viable or even preferable for some applications.  For given $\theta$, we decompose the $n_{y}\times n_{y}$ covariance matrix $\Sigma_{y}$ as,
\begin{equation}\label{eq:covparam}
	\mathsf{T}(\theta)\,\Sigma_{y}(\theta)\, \mathsf{T}^{T}(\theta) = \mathsf{D}(\theta) \quad \text{or} \quad \Sigma_{y}^{-1} = \mathsf{T}^{T}\,\mathsf{D}^{-1}\, \mathsf{T},
\end{equation}
where $\mathsf{D}$ is a diagonal matrix of strictly positive ``variances'' and $\mathsf{T}$ is a lower triangular matrix with ones on the diagonal and unconstrained off-diagonal elements 
\[
\varphi_{ij}\equiv - \mathsf{T}_{ij}\qquad 2\le i\le n_{y}, \quad j = 1,\dots,i-1.
\]
The fact that the $\varphi_{ij}$ are unconstrained makes this a computationally convenient parameterization.  In addition, because the decomposition of the inverse covariance is quadratic in the $\varphi_{ij}$, the conjugate prior for the $\varphi_{ij}$ is a Gaussian.  This will be very convenient when we specify our interpolation method below.  A conjugate Gaussian prior allows us to impose prior structure on $\Sigma_{y}$ via the mean and covariance of $\varphi_{ij}$.  Considered as a single column vector for given $\theta$,
\begin{equation}\label{eq:phiprior}
	\varphi \sim \text{N}\left(\bar{\varphi},C_{\varphi}\right).
\end{equation}
Note that $\bar{\varphi}$ and $C_{\varphi}$ are independent of $\theta$ so that we can ``shrink'' the covariance matrix estimates towards a parameter-independent $\mathsf{T}$.  The prior mean, $\bar{\varphi}$, can be constructed from the generalized Cholesky decomposition of the the average sample covariance matrix from the simulation runs, 
\[
	\hat{S}_{y} \equiv \frac{1}{n_{d}} \sum_{i=1}^{n_{d}} \tilde{S}_{y,i},
\]
where $\tilde{S}_{y,i}$ is the sample covariance matrix at the $i$th design point.  If there are not enough simulation runs to get good estimates of $\tilde{S}_{y}$, the sample covariance of the combined simulation runs could be used instead,
\[
	\tilde{S}_{\text{design}} = \frac{1}{m} \sum_{i=1}^{m} (y_{i}-\mu_{i})(y_{i}-\mu_{i})^{T}.
\]
The prior covariance, $C_{\varphi}$, could be diagonal with separate variances for each $\varphi_{ij}$ when little prior knowledge about the structure of the $\Sigma_{y,i}$ is known.  A slightly more informative prior is the generalized inverse Wishart prior~\cite{brown94} with scale matrix $\hat{S}_{y}$ or $\tilde{S}_{\text{design}}$.  In this case, $C_{\varphi}$ takes a block diagonal structure as described in Eqns.~(12-17) of Ref.~\cite{daniels02}.

The number of components to model can be reduced by expanding $\varphi$ in a set of basis functions (or covariates) so that 
\begin{equation}\label{eq:phidecomp}
	\varphi_{ij} = \sum_{k=1}^{p_{\varphi}} Z_{ij}^{k}\gamma_{k}\quad 
	p_{\varphi} \le \half n_{y}(n_{y}-1).
\end{equation}
This decomposition preserves the quadratic dependence of the log-likelihood on the variables, so a conjugate Gaussian prior can be specified on the $\gamma_{k}$.

In analogy with Eqn.~(\ref{eq:gpdists}), we model the individual $\varphi_{i}$'s as GPs, with the same covariance structure as in Eqn.~(\ref{eq:GPcov}),
\begin{eqnarray}
	\varphi_{i}(\theta) \sim &\text{GP}\left(\phibar_{i},\Sigma_{\varphi}(\theta;\lambda_{\varphi,i},\brho_{\varphi,i})\right) 
	& i=1,\dots,\frac{n_{y}(n_{y}-1)}{2}.
\end{eqnarray}
Note that if the decomposition in Eqn.~(\ref{eq:phidecomp}) is used, then $\gamma_{i}$ can be substituted for $\varphi_{i}$ above.  Restricted to the design points, the prior for $\varphi$ becomes,
\begin{eqnarray}
	\phis &\sim& {\rm N}\left(\phibar,\Sigma_{\varphi}^{*}(\blambda_{\varphi},
	\brho_{\varphi})\right),
\end{eqnarray}
where $\phis$ has length $\half n_{y}(n_{y}-1)n_{d}$.

The sampling distribution for $\phis$ is just the product of the GP prior on $\phis$ times the prior in Eqn.~(\ref{eq:phiprior}), which gives an unnormalized Gaussian distribution for $\phis$,
\begin{eqnarray}\label{eq:phisamp}
	&&\pi(\phis|\blambda_{\varphi},\brho_{\varphi}) = 
	\left|C_{\varphi}\otimes\ident_{n_{d}}\right|^{-1/2}\,
	\left|\Sigma_{\varphi}^{*}\right|^{-1/2}
	\\
	&&\times 
	\exp\left\{ (\phis-\phibar)^{T}
	\left[ \Sigma_{\varphi}^{*-1}+
	\left(C_{\varphi}\otimes\ident_{n_{d}}\right)^{-1} \right]
	(\phis-\phibar)\right\}.\nonumber
\end{eqnarray}

\section{\label{sc:emlike}Full emulator likelihood}
The expression for the joint likelihood of the data and simulation runs in Eqn.~(\ref{eq:finallike}) can be simplified further by performing the integral over $\mus$.   
If we collect all the $\mus$-dependent terms in the integrand of Eqn.~(\ref{eq:finallike}), we can write the conditional distribution for $\mus$ as,
\begin{eqnarray}\label{eq:musintegrand}
	\pi(\mus|\Ys,\Ds,\phis,\hat{w}_{y},v_{0},\varphi_{0},\bomega) = 
	L(\Ys|\mus,\Ds,\phis)\cdot \pi(\hat{w}|\hat{w}_{y},v_{0},\varphi_{0},\theta_{0},\bomega)\cdot \rprior(\mus|\lemu).
\end{eqnarray}
For this section, we have included the covariance matrix parameterization from Appendix~\ref{sc:covparam}, which accounts for the extra $\varphi$ factors above.  

Using Eqn.~(\ref{eq:mvmodel}), we can write an explicit expression for the likelihood of the simulation design outputs,
\begin{eqnarray}\label{eq:simlikeexp}
	\ln\left(L(\Ys|\mus,\Ds,\phis)\right) &=& -\frac{\mu_{s}^{2}}{2}
	\sum_{i=1}^{n_{d}}\sum_{j=1}^{n_{r_{i}}} \left(\Yst_{ij}-\must_{i}\right)^{T} 
	\Sigma_{y}^{-1}\left[\Ds_{i},\phis_{i}\right] \left(\Yst_{ij}-\must_{i}\right)\\
	 &-&\half\sum_{i=1}^{n_{d}}\,n_{r_{i}}
	\sum_{j=1}^{n_{y}}\left(D_{s}\ln\left(\Dst_{ij}\right)+D_{c,j}\right) 
	+ \text{constant}
	\nonumber\\
	&=& -\frac{\mu_{s}^{2}}{2} \sum_{i=1}^{n_{d}}\left[
	\left(\must_{i} - \bar{Y}^{*}_{i}\right)^{T}n_{r_{i}}\Sigma_{y,i}^{-1}
	\left(\must_{i} - \bar{Y}^{*}_{i}\right)
	- \bar{Y}^{*T}_{i}n_{r_{i}}\Sigma_{y,i}^{-1}\bar{Y}^{*}_{i} 
	+ \sum_{j=1}^{n_{r_{i}}}\tilde{Y}^{*T}_{ij}\Sigma_{y,i}^{-1}\Yst_{ij}
	\right],\nonumber
\end{eqnarray}
where $\tilde{Y}_{i}=\frac{1}{n_{r_{i}}} \sum_{j=1}^{n_{r_{i}}} Y_{ij}$ is the sample mean at each design point.

\begin{eqnarray}
	-2\,\ln\left(\pi(\hat{w}|\hat{w}_{y},v_{0},
	\varphi_{0},\theta_{0},\bomega)\right) 
	&=& \left(\hat{w}-\Swwy\Swy^{-1}\hat{w}_{y}\right)^{T} 
	\left(\Swb-\Swwy\Swy^{-1}\Swwy^{T}\right)^{-1}
	\left(\hat{w}-\Swwy\Swy^{-1}\hat{w}_{y}\right)
	\nonumber\\
	&& + 
	\ln\left| \Swb-\Swwy\Swy^{-1}\Swwy^{T} \right|
	+ \text{constant}
	\nonumber\\
	&\equiv& \left(\hat{w}-z_{y}\right)^{T} S_{w_{y}}^{-1}
	\left(\hat{w}-z_{y}\right) + \ln\left|S_{w_{y}}\right| + \text{constant},
\end{eqnarray}
where $\Swy\equiv \Slw + \left(\Phi_{\mu}^{T}W_{y}\Phi_{\mu}\right)^{-1}$ and $\Swb\equiv \lemu^{-1}\ident_{n_{d}p_{\mu}} + \Sigma_{\hat{w}}$,
\begin{eqnarray}
	-2\,\ln\left(\rprior(\mus|\lemu)\right) 
	&=& \lemu \mu^{*T}\left(\ident_{q}-\Phi_{\mu}\Phi_{\mu}^{T}\right)\mus 
	-n_{d}(n_{y}-p_{\mu})\ln\left(\lemu\right) + \text{constant}.
\end{eqnarray}

The final expression for the joint likelihood of the data and simulation runs becomes,
\begin{align}\label{eq:finallike2}
	L(y,\Ys|\theta_{0},\bomega) &= \int\int d\Ds\,d\phis\, \int dv_{0}\,d\phi_{0}\,\,
	L(\Ys|\Ds,\phis,\hat{w}_{y},v_{0},\varphi_{0},\theta_{0},\bomega)\\
	&\quad\times\pi(\hat{w}_{y}|v_{0},\varphi_{0})\cdot
	\rprior(y|v_{0},\phi_{0})\cdot
	\pi(v_{0},\hat{v}|\theta_{0},\bomega)\cdot\rprior(\Ds|\led)\cdot
	\pi(\varphi_{0},\phis|\theta_{0},\bomega),\nonumber
\end{align}
with,
\begin{eqnarray}
	-2\ln\left(L(\Ys|\Ds,\phis,\hat{w}_{y},v_{0},\phi_{0},\theta_{0},\bomega)\right)
	&=& 
	\ln\left|	C_{y} + \Sigma_{\mu_{p}}\right|
	+(x^{*}-z)^{T}\left(C_{y}+\Sigma_{\mu_{p}}\right)^{-1}(x^{*}-z)
	\nonumber\\
	&&+\, \mu_{s}^{2}\sum_{i=1}^{n_{d}}\left[
	\sum_{j=1}^{n_{r_{i}}}\left(\tilde{Y}^{*T}_{ij}\Sigma_{y,i}^{-1}\Yst_{ij}\right)
	-\bar{Y}^{*T}_{i}n_{r_{i}}\Sigma_{y,i}^{-1}\bar{Y}^{*}_{i} \right],
	\nonumber\\
	-2\ln\left(\pi(\hat{w}_{y}|v_{0},\phi_{0})\right) &=& 
	\ln\left|\bar{\Sigma}_{w_{y}}\right| + 
	\hat{w}_{y}^{T}\bar{\Sigma}_{w_{y}}^{-1}\hat{w}_{y} 
	+ \text{constant},\nonumber\\
	-2\ln\left(\rprior(y|v_{0},\phi_{0})\right) &=&
	-\ln\left|W_{y}\right|
	-\ln\left|\Phi_{\mu}^{T}W_{y}\Phi_{\mu}\right| + 
	\left(\tilde{y}-\Phi_{\mu}\hat{w}_{y}\right)^{T}W_{y}
	\left(\tilde{y}-\Phi_{\mu}\hat{w}_{y}\right)\nonumber\\
	&+& \text{constant},\nonumber\\
	-2\ln\left(\pi(v_{0},\hat{v}|\theta_{0},\bomega)\right) &=&
	\ln\left|S_{\hat{v}}\right| + 
	\left(v_{0}-\Sigma_{\hat{v}v}^{T}\bar{\Sigma}_{\hat{v}}^{-1}\hat{v}\right)^{T}
	S_{\hat{v}}^{-1}
	\left(v_{0}-\Sigma_{\hat{v}v}^{T}\bar{\Sigma}_{\hat{v}}^{-1}\hat{v}\right)
	\notag\\
	&&\qquad + \ln\left|\bar{\Sigma}_{\hat{v}}\right| 
	+ \hat{v}^{T}\bar{\Sigma}_{\hat{v}}^{-1}\hat{v},
	\notag\\
	-2\ln\left(\rprior(\Ds|\led)\right) &=& 
	-\,n_{d}(n_{y}-p_{D})\ln(\led)
	+ \led \ln\left(\tilde{D}^{*T}\right) \left(I_{q}-\Phi_{D}^{*}\Phi_{D}^{*T}
	\right)\ln\left(\Dst\right),\notag\\
	-2\ln\left(\pi(\phis,\varphi_{0}|\theta_{0},\bomega)\right) &=&
	\text{extension of Eqn.~(\ref{eq:phisamp})},\notag
\end{eqnarray}
 $S_{\hat{v}}$ is the Schur complement of $\bar{\Sigma}_{\hat{v}}$ in the joint covariance for $\hat{v},v_{0}$,
\begin{eqnarray}
	C_{y,i}^{-1}\equiv \mu_{s}^{2}\,n_{r_{i}}
	\Phi_{\mu}^{T}\Sigma_{y,i}^{-1}\Phi_{\mu},
	\qquad
	\Sigma_{\mu_{p}} \equiv\left(
	\begin{array}{cc}
		S_{w_{y}} & 0 \\
		0 & \lemu^{-1}\ident_{n_{d}(n_{y}-p_{\mu})}
	\end{array}
	\right),
\end{eqnarray}
$x^{*}_{i}\equiv\Phi_{\mu}^{T}\bar{Y}^{*}_{i}$ for $i=1\dots n_{d}$, and,
\[
	z \equiv \left(
	\begin{array}{c}
		\Swwy\Swy^{-1}\hat{w}_{y}\\
		0
	\end{array}
	\right) \qquad (\text{dimensions:}\, n_{y}\times 1).
\]

\section{\label{sc:proposal} Proposal distributions for Metropolis MCMC updates}
We use the prior on $v(\theta)$ as a proposal distribution for the Metropolis updates in performing the Monte Carlo integral in Eqns.~(\ref{eq:finallike}) or (\ref{eq:finallike2}).  We rewrite the joint prior in Eqn.~(\ref{eq:vjointlike}) as
\begin{equation}
	\pi(v,\hat{v}|\theta_0,\led,\lv,\rv) = \pi(v|\hat{v},\theta_{0},\led,\lv,\rv)\cdot 
	\pi(\hat{v}|\led,\lv,\rv),
\end{equation}
where, using Eqn.~(\ref{eq:vjointlike}) and the conditional Normal rule,
\begin{eqnarray}
	v|\hat{v},\theta_{0},\led,\lv,\rv &\sim& {\rm N}\left( \Sigma_{\hat{v}v}^{T}
	\bar{\Sigma}_{\hat{v}}^{-1}\hat{v}, \Sigma_{\lambda_{v}} - 
	\Sigma_{\hat{v}v}^{T}\bar{\Sigma}_{\hat{v}}^{-1}\Sigma_{\hat{v}v}\right), \nonumber\\
	\hat{v}|\led,\lv,\rv &\sim& {\rm N}\left(0,\bar{\Sigma}_{\hat{v}}\right),
\end{eqnarray}
and we have defined the shorthand, $\bar{\Sigma}_{\hat{v}}\equiv \led^{-1}\ident + \Sigma_{\hat{v}}$.  
 
 \section{\label{sc:priors} Priors for the emulator hyperparameters}
 The full joint posterior of the cosmological and GP parameters is,
 \begin{eqnarray}
 	\pi\left( \theta_{0},\lemu,\lw,\rw,\led,\lv,\rv| y,\tilde{\mu}^{*},\tilde{D}^{*}\right)
	&\propto& L\left( y,\tilde{\mu}^{*},\tilde{D}^{*}| 
	\theta_{0},\lemu,\lw,\rw,\led,\lv,\rv\right) \nonumber\\*
	&\times& \pi\left(\lemu\right)\pi(\lw)\pi(\rw)
	\pi\left(\led\right)\pi(\lv)\pi(\rv)\pi(\theta_{0}),\nonumber
 \end{eqnarray}
 where the likelihood is given in Eqns.~(\ref{eq:finallike2}) or (\ref{eq:simpemlike}), and copying~\cite{habib07},
 \begin{eqnarray}
 	\pi\left(\lemu\right) &\propto& \lemu^{a_{\mu}-1} e^{-b_{\mu}\lemu}, 
	\nonumber\\
	\pi(\lw) &\propto& \prod_{i=1}^{p_{\mu}} \lambda_{w_i}^{a_w-1} 
	e^{-b_w\lambda_{w_i}}, 
	\nonumber\\
	\pi(\rw) &\propto& \prod_{i=1}^{p_{\mu}}
	\prod_{j=1}^{p_{\theta}} \rho_{w_{ij}}^{a_{\rho_w}-1} 
	(1-\rho_{w_{ij}})^{b_{\rho_w}-1}, 
	\nonumber\\
	\pi\left(\led\right) &\propto& \led^{a_{D}-1} e^{-b_{D}\led}, 
	\nonumber\\
	\pi(\lv) &\propto& \prod_{i=1}^{p_{D}} 
	\lambda_{v_i}^{a_v-1} e^{-b_v\lambda_{v_i}},
	\nonumber\\
	\pi(\rv) &\propto& \prod_{i=1}^{p_{D}} \prod_{j=1}^{p_{\theta}}
	\rho_{v_{ij}}^{a_{\rho_v}-1} (1-\rho_{v_{ij}})^{b_{\rho_v}-1},\,\text{and}
	\nonumber\\
	\pi(\theta_{0}) &=& {\rm uniform}(0,1)\, 
	\text{for each}\, \theta_{0,i}\quad
	i=1,\dots,p_{\theta},
 \end{eqnarray}
 with
$a_{\mu}=a_{D}=1$, $b_{\mu}=b_{D}=0.0001$, $a_{w}=a_{v}=5$, $b_{w}=b_{v}=5$, 
 $a_{\rho_{w}}=a_{\rho_{v}}=1$, $b_{\rho_{w}}=b_{\rho_{v}}=0.2$.

\section{\label{sc:covmatexpressions}Explicit expressions for covariance matrices}
To evaluate the distributions in Eqns.~(\ref{eq:wjointlike}) and (\ref{eq:vjointlike}), we use,
 \begin{eqnarray}
	\Sigma_{\lambda_{w}} &=& {\rm diag}\left(\lambda_{w_{i}}^{-1}\right) \qquad
	(\text{dimensions:}\,p_{\mu}\times p_{\mu}), \nonumber\\
	\Sigma_{\lambda_{v}} &=& {\rm diag}\left(\lambda_{v_{i}}^{-1}\right) \qquad
	(\text{dimensions:}\,p_{D}\times p_{D}),\nonumber
\end{eqnarray}
\begin{equation}
	\Sigma_{\hat{w}} = \left(
	\begin{array}{ccc}
	\Lambda_{w_{1}} & 0 & 0 \\
	0 & \ddots & 0 \\
	0 & 0 & \Lambda_{w_{p_{\mu}}}
	\end{array}\right) \qquad (\text{dimensions:}\,(n_{d}\,p_{\mu})\times(n_{d}\,p_{\mu})),
\end{equation}
following Eqn.~(15) of Ref.~\cite{habib07}, with,
\begin{equation}
	\Lambda_{w_{i}} = \lambda_{w_{i}}^{-1} R(\theta^{*};\rho_{w_{i}}) \qquad
	(\text{dimensions:}\,n_{d}\times n_{d}),
\end{equation}
\begin{equation}
	\Sigma_{\hat{w}\, w_{y}} = \left(
	\begin{array}{ccc}
	\lambda_{w_{1}}^{-1}R(\theta,\theta^{*};\rho_{w_{1}}) & 0 & 0 \\
	0 & \ddots & 0 \\
	0 & 0 & \lambda_{w_{p_{\mu}}}^{-1}R(\theta,\theta^{*};\rho_{w_{p_{\mu}}})
	\end{array}\right)
	\qquad (\text{dimensions:}\,(n_{d}\,p_{\mu})\times p_{\mu}),
\end{equation}
and $R(\theta,\theta^{*};\rho_{w_{i}})$ is a $n_{d}\times 1$ correlation sub-matrix.  Analogous expressions hold for $\Sigma_{\hat{v}}$ and $\Sigma{\hat{v}v}$.

We invert the full covariance matrices in  Eqns.~(\ref{eq:wjointlike}) and (\ref{eq:vjointlike}) using the block-inverse formula,
\begin{equation}
	\left(\begin{array}{cc}
	A & B \\
	B^{T} & D
	\end{array}\right)^{-1}
	= \left(\begin{array}{cc}
	(A-BD^{-1}B^{T})^{-1} & -A^{-1}B(D-B^{T}A^{-1}B)^{-1} \\
	-(D-B^{T}A^{-1}B)^{-1}B^{T}A^{-1} & (D-B^{T}A^{-1}B)^{-1}
	\end{array}\right).
\end{equation}
For the $w$ likelihood, $A=\lambda_{\epsilon_{\mu}}^{-1}\ident_{n_{d}\,p_{\mu}} + \Sigma_{\hat{w}}$, $B = \Sigma_{\hat{w}\,w_{y}}$, $D = \Sigma_{\lambda_{w}} + (\Phi_{\mu}^TW_{y}\Phi_{\mu})^{-1}$.  For the $v$ likelihood, $A=\lambda_{\epsilon_{D}}^{-1} \ident_{n_{d}p_{D}} + \Sigma_{\hat{v}}$, $B=\Sigma_{\hat{v}\,v}$, $D=\Sigma_{\lambda_{v}}$.

\end{widetext}

\begin{acknowledgments}
We would like to thank Roman Scoccimarro, Benjamin Wandelt, and Martin White for useful discussions.  S.H., K.H., and D.H. acknowledge support from the LANL
LDRD program.  The computations for this paper were carried out using the Scythe Statistical Library~\cite{scythestatlib}.  This work was supported in part at UC Davis by NSF Grant AST-0709498.
\end{acknowledgments}

\bibliography{compexp}
\end{document}